\documentclass[twocolumn,prl,floatfix,showpacs,preprintnumbers,amsmath,amssymb]{revtex4}

\usepackage{color}
\usepackage{graphicx}
\usepackage{dcolumn}    
\usepackage{multirow}
\usepackage{bm}         
\usepackage{sidecap}
\newcommand{\bsigma}{\mbox{\boldmath$\sigma$}}
\newcommand{\btau}{\mbox{\boldmath$\tau$}}
\begin{document}
%

\title{A new Skyrme interaction with improved spin-isospin properties}

\author{X. Roca-Maza\textsuperscript{1}}
\author{G. Col\`o\textsuperscript{1,2}}
\author{H. Sagawa\textsuperscript{3,4}}

\affiliation{\textsuperscript{1} INFN, sezione di Milano, via Celoria 16, 
             I-20133 Milano, Italy\\
             \textsuperscript{2} Dipartimento di Fisica, Universit\`a degli 
             Studi di Milano, via Celoria 16, I-20133 Milano, Italy\\
             \textsuperscript{3} Center for Mathematics and Physics, University 
             of Aizu, Aizu-Wakamatsu, Fukushima 965-8560, Japan\\
           \textsuperscript{4} Nishina Center, Wako,  Saitama  351-0198, Japan}

\date{\today} 

%
\begin{abstract}
A correct determination of the spin-isospin properties of the nuclear effective interaction should lead, among other improvements, to an accurate description of the Gamow-Teller Resonance (GTR). These nuclear excitations impact on a variety of physical processes: from the response in charge-exchange reactions of nuclei naturally present in the Earth, to the description of the stellar nucleosynthesis, and of the pre-supernova explosion core-collapse evolution of massive stars in the Universe. A reliable description of the GTR provides also stringent tests for neutrinoless double-$\beta$ decay calculations. We present a new Skyrme interaction as accurate as previous forces in the description of finite nuclei and of uniform matter properties around saturation density, and that account well for the GTR in ${}^{48}$Ca, ${}^{90}$Zr and ${}^{208}$Pb, the Isobaric Analog Resonance and the Spin Dipole Resonance in ${}^{90}$Zr and ${}^{208}$Pb. 
\end{abstract}

\pacs{21.60.Jz, 24.30.Cz}



\maketitle

The Skyrme Hartree-Fock (HF) approach is one of the successful techniques for the study of the ground state properties of nuclei and, if supplemented by a proper description of nuclear superfluidity (e.g., within the Hartree-Fock-Bogoliubov scheme), it can be applied throughout the whole periodic table \cite{ston07}. The small amplitude limit of time-dependent HF calculations, or Random Phase Approximation (RPA), has allowed to describe many kinds of nuclear collective motion \cite{paar07}. The versatility of the Skyrme ansatz allows its use in more elaborated theoretical frameworks that include higher-order nuclear correlations - like the Generator Coordinate Method \cite{ref_GCM}, or the Particle-Vibration Coupling approach \cite{ref_PVC}.  

Despite the existence of drawbacks and open issues, the Skyrme-HF approach enables an effective description of the nuclear many-body problem in terms of a local energy density functional. Problems concerning specific terms of this functional need to be understood and eventually solved (cf. also Ref. \cite{erl11}). One of these problems, and the focus of the present work, is the accurate determination of the spin-isospin properties of the Skyrme effective interaction and of the associated functional. Such a determination should lead to accurate predictions of the properties of GTR, that are among the clearest manifestations of nuclear collective motion \cite{ost92}. Gamow-Teller (GT) transitions determine weak-interaction rates between $fp$-shell nuclei that play an essential role in the core-collapse dynamics of massive stars leading to supernova explosion \cite{beth90,lang08} (in this neutron-rich environment, neutrino-induced nucleosynthesis may take place via GT processes \cite{byel07}). Accurate GT matrix elements are necessary for the study of double-$\beta$ decay \cite{avig08}, and may be useful in the calibration of detectors aiming to measure electron-neutrinos coming from the Sun \cite{lass02}.  

The earliest attempt to give a quantitative description of the GTR data was provided by the Skyrme SGII interaction \cite{giai81}. However, two component spin-orbit  contributions to the nuclear Hamiltonian density -- Eq. (6.1) in Ref.~\cite{chab98} -- were not introduced. Later on, using the full Skyrme Hamiltonian density, a functional suitable for the predictions of finite nuclei and charge-exchange resonances was proposed: namely SkO' \cite{rein99}. Relativistic mean-field and relativistic HF calculations of the GTR have also become available meanwhile \cite{paar04,lian08}. In a recent and detailed study on the GTR and the spin-isospin Landau-Migdal parameter ($G_0^\prime$) using several Skyrme sets \cite{bend02}, Bender {\it et al.} have concluded that this spin-isospin coupling is not the only important quantity in determining the strength and excitation energy of the GTR in nuclei. Actually, the authors state that spin-orbit splittings together with the residual spin-isospin interaction influences the above mentioned quantities.

The GT transition strength ($R_{{\rm GT}^\pm}$) is mediated by the operator $\sum_{i=1}^A \bsigma(i)\tau_\pm(i)$, where $A$ is the mass number and $\bsigma$ and $\btau$ are the spin and isospin Pauli matrices, respectively. The contributions to $R_{{\rm GT}^\pm}$ come from nucleon transitions that change the spin and isospin of the parent quantum state and the residual interaction between them is repulsive. The dominant transitions will be those between spin-orbit partner levels. In this respect, most of the Skyrme interactions overestimate the experimental spin-orbit splittings in heavy nuclei \cite{bend99}. 

Experimentally, the GTR exhausts only about 60-70\% of the well known Ikeda Sum Rule (ISR) given by $\int [R_{{\rm GT}^-}(E) - R_{{\rm GT}^+}(E)]{\rm d}E=3(N-Z)$. To explain this well-known quenching problem, it has been proposed that the effects of the second-order configuration mixing, namely 2-particle 2-hole (2p-2h) correlations, or of the coupling with the  $\Delta -$hole   excitation, have to be taken into account. The experimental analysis of $^{90}$Zr \cite{waka97} seems to indicate that most of the quenching (around 2/3) has to be attributed to 2p-2h coupling while the role played by the $\Delta$ isobar is much smaller. 

In our work, we present a new non-relativistic functional of the Skyrme type that include the central tensor terms ($J^2$-terms) and two spin-orbit parameters. It is named SAMi for {\it S}-kyrme {\it A}-izu {\it Mi}-lano. The new functional is as accurate as previous Skyrme models in the description of uniform nuclear matter properties around saturation and of masses and charge radii of double-magic nuclei. It is also precise in the description of the Giant Monopole Resonance (GMR), the Giant Dipole Resonance (GDR) in ${}^{208}$Pb and the GTR, the Isobaric Analog Resonance (IAR) and the Spin Dipole Resonance (SDR) in medium and heavy mass nuclei.
\begin{table}[t]
\begin{center}
\caption{Data and {\it pseudo}-data $\mathcal{O}_i$, adopted errors for the fit $\Delta\mathcal{O}_i$, as well as partial and total number of data points and contributions to the $\chi^2$.}
\begin{tabular}{llrrr}
\hline\hline
$\mathcal{O}_i$ & $\Delta\mathcal{O}_i$&  $\chi^2_{\rm partial}$&$n_{\rm data}$& Ref.\\
\hline
$B$               & 1.00 MeV                 & 32.45 & 5 & \cite{audi03}\\ 
$r_{c}$            & 0.01 fm                  & 13.38 & 4 & \cite{ange04} \\
$\Delta E_{\rm SO}$ & 0.04$\times\mathcal{O}_i$& 19.02 & 2 & \cite{zale08} \\
$e_n(\rho)$       & 0.20$\times\mathcal{O}_i$&  12.60 &11 & \cite{wiri88} \\
\hline
$\chi^2$          &                          & 77.45&/ 22 &= 3.52          \\
\hline\hline
\end{tabular}
\label{data}
\end{center}
\vspace{-0.5cm}
\end{table}

To this end, we have carefully chosen the set of fitted data and $pseudo$-data inspired by the protocol used to build SLy interactions \cite{chab98}: (i) the binding energies ($B$) of ${}^{40,48}$Ca, ${}^{90}$Zr, ${}^{132}$Sn and ${}^{208}$Pb and the charge radii ($r_c$) of ${}^{40,48}$Ca, ${}^{90}$Zr and ${}^{208}$Pb which allow us to determine the saturation energy ($e_\infty$), density ($\rho_\infty$) and, to a good approximation the incompressibility ($K_\infty$) of symmetric nuclear matter; (ii) the spin-orbit splittings ($\Delta E_{\rm SO}$) of the 1$g$ and 2$f$ proton levels in ${}^{90}$Zr and ${}^{208}$Pb, respectively, that are well determined due to the flexibility of our two-component spin-orbit potential; and (iii) the Landau-Migdal parameters $G_0$ and $G_0^\prime$ -- associated with the spin and spin-isospin particle-hole (p-h) interaction (see their definition in Ref.~\cite{gang10}) that are fixed at the values 0.15 and 0.35, respectively, at saturation density. This features allow the new SAMi interaction to give an adequate description of spin-isospin resonances. In the literature, an empirical determination of the Landau-Migdal parameters can be found in Ref.~\cite{waka07} but we do not use such values as pseudo-data in our fit. The reason is that the extraction of these values is based on single-particle energies obtained with a Woods-Saxon potential. In our case, we use HF energies associated with a different effective mass. However, we took inspiration form the empirical indications that suggest $G_0^\prime > G_0 > 0$. This is not a very common feature within available Skyrme forces (see Fig. 1 in Ref.~\cite{gang10}). Therefore, we imposed that $G_0$ is larger than 0 and $G_0^\prime$ is larger than 0.25, and we have tried to explore the  optimum values that do not spoil the global fit; finally, (iv) {\it pseudo}-data corresponding to more fundamental microscopic calculations of the energy per particle of uniform neutron matter ($e_{n}$) at baryon density $\rho$ between $0.07$ fm${}^{-3}$ and $0.4$ fm${}^{-3}$ that have been helpful in driving the magnitude ($J$) and slope ($L$) of the nuclear symmetry energy at normal densities towards reasonable values. Table \ref{data} provides references for these data and {\it pseudo}-data with the corresponding adopted errors, partial contributions to the $\chi^2$, and the number of data points ($n_{\rm data}$) used in the fit. The main differences between the SLy \cite{chab98} and the present protocol are the fitting of the above mentioned spin-orbit splittings, the fact that we fix the spin and spin-isospin Landau-Migdal parameters,  and the larger adopted errors for the equation of state of pure neutron matter. This protocol is justified by the fact that {\it pseudo}-data are used as a guide and, therefore, it should not impact on the fitted interaction more than experimental data.
\begin{table}[b]
\vspace{-0.5cm}
\begin{center}
\caption{SAMi parameter set and saturation properties (see text) with the estimated standard deviations \cite{bevi92} inside parenthesis (referred to the last digits).}
\begin{tabular}{lrllrl}
\hline\hline
            &value($\sigma$)   &                       &   & value($\sigma$) &\\
\hline
 $t_0$      &$-$1877.75(75)    &MeV fm${}^{3}$         &$\rho_\infty$  &     0.159(1)& fm${}^{-3}$\\
 $t_1$      &    475.6(1.4)    &MeV fm${}^{5}$         &$e_\infty$     &$-$15.93(9)& MeV        \\
 $t_2$      &  $-$85.2(1.0)    &MeV fm${}^{5}$         &$m^*_{\rm IS}$ &    0.6752(3)  &            \\
 $t_3$      &  10219.6(7.6)    &MeV fm${}^{3+3\alpha}$   &$m^*_{\rm IV}$ &    0.664(13)  &            \\
 $x_0$      &      0.320(16)   &                       &$J        $   &   28(1)  & MeV        \\
 $x_1$      &   $-$0.532(70)   &                       &$L        $   &   44(7)  & MeV        \\
 $x_2$      &   $-$0.014(15)   &                       &$K_\infty $    &  245(1)      & MeV        \\
 $x_3$      &      0.688(30)   &                       &$G_0      $    &    0.15    & (fixed)      \\
 $W_0$      &    137(11)       &                       &$G_0^\prime $   &    0.35    & (fixed)      \\
 $W_0^\prime$ &   42(22)       &                        &               &             &            \\
 $\alpha$   &      0.25614(37) &                       &               &             &            \\
\hline\hline
\end{tabular}
\label{parprops}
\end{center}
\end{table}
The minimization of the $\chi^2$ has been performed by means of a variable metric method included in the {\sc MINUIT} package of Ref.~\cite{jame96}.  

\begin{figure}[t]
\includegraphics[width=0.9\linewidth,clip=true]{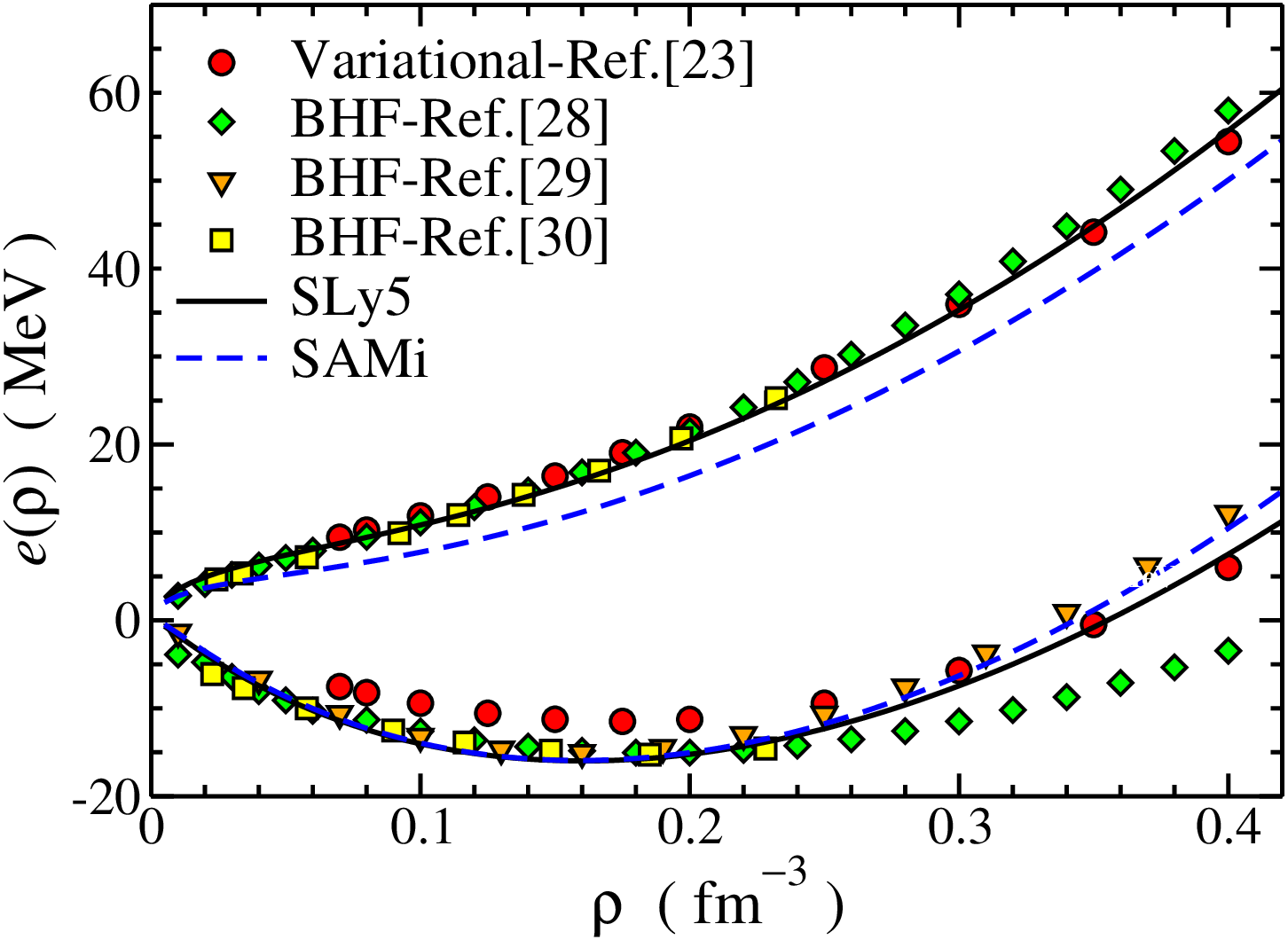}
\caption{Neutron and symmetric matter EoS as predicted by the HF SAMi (dashed line) and SLy5 (solid line) interactions and by the benchmark microscopic calculations of Ref.~\cite{wiri88} (circles). State-of-the-art BHF calculations are shown by diamonds \cite{vida12}, triangles \cite{li08} and squares \cite{bald04}.}
\label{eos}
\vspace{-0.5cm}
\end{figure}

The parameters and saturation properties of the new interaction are shown in Table \ref{parprops}. The estimation of the standard deviation \cite{bevi92} associated to each of them is also displayed. In what follows, the new SAMi functional is compared to available experimental data and other theoretical predictions for ground and excited state properties. First of all, we show in Fig.~\ref{eos} the results for the symmetric and pure neutron matter Equations of State (EoS) as predicted by the benchmark microscopic calculations used in the fit \cite{wiri88}, three state-of-the-art Brueckner-Hartree-Fock (BHF) calculations \cite{bald04,vida12,li08}, the SAMi functional,  and SLy5 \cite{chab98} -- also fitted to reproduce the neutron matter EoS of Ref.~\cite{wiri88}. The agreement of the SAMi functional with these calculations of nuclear matter based on realistic nucleon-nucleon (NN) forces is remarkable. The deviation of the SAMi EoS of pure neutron matter from the fitted microscopic curve (red circles) is essentially due to the relatively large error (20\%) adopted in the $\chi^2$ definition in order not to spoil the additional constraints set on the isovector channel of the effective interaction. In addition, we have checked that the SAMi EoS is stable against spin and spin-isospin instabilities \cite{gang10} up to a baryon density of 4.1$\rho_\infty$ and 5.3$\rho_\infty$, respectively, i.e., well above the region important for the description of finite nuclei and enough for the study of uniform neutron-rich matter in neutron stars. Furthermore, we are aware that particle-number projection techniques lead to instabilities when functionals with non-integer power of the density are employed \cite{dugu09}. At the same time, the adopted density dependence ($\rho^\alpha$ with $\alpha$ smaller than one) seems to be the only way to have reasonable values of the nuclear incompressibility and of the GMR energies within the Skyrme functional. As a future perspective, all practitioners of local, Skyrme type, EDFs may need to deal with the problem of reproducing reasonable monopole energies on the one side and making particle number restoration doable on the other side. This is beyond the purpose of the current work.   

\begin{SCfigure*}
\includegraphics[width=0.7\linewidth,clip=true]{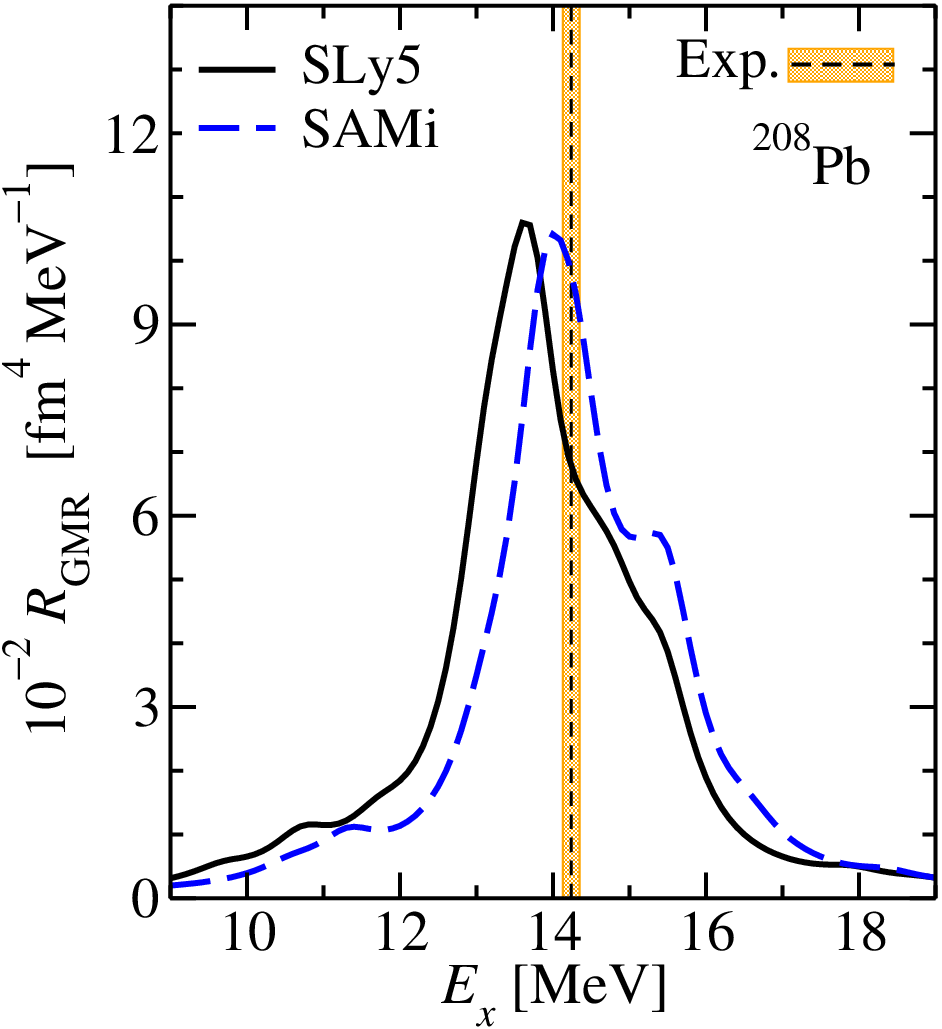}  
\includegraphics[width=0.7\linewidth,clip=true]{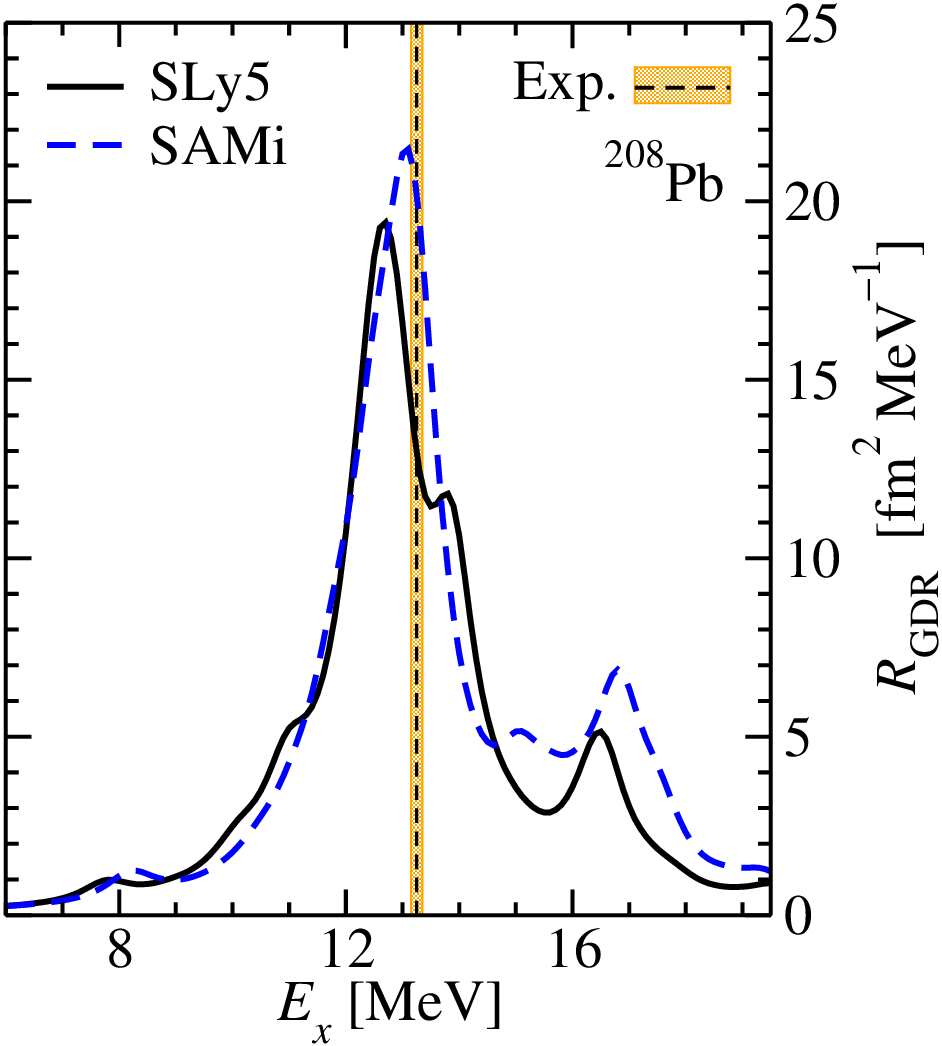}
\caption{Strength function at the relevant excitation energies in 
${}^{208}$Pb as predicted by SLy5~\cite{chab98} and the SAMi interaction for 
GMR (left panel) and GDR (right panel). A Lorentzian smearing 
parameter equal to 1 MeV is used. Experimental data for the 
centroid energies is also shown: 
$E_{\rm c}({\rm GMR})=14.24\pm0.11$ MeV~\cite{youn99} and 
$E_{\rm c}({\rm GDR})=13.25\pm 0.10$ MeV \cite{ryez02}. 
\vspace{1.5cm}}
\label{gmr-gdr}
\vspace{-0.5cm}
\end{SCfigure*}

\begin{table}[b]
\vspace{-0.5cm}
\begin{center}
\caption{Experimental data \cite{audi03,ange04,zale08} and SAMi 
results for the binding energies ($B$), charge radii ($r_c$), and 
proton spin-orbit splittings [$\Delta E_{\rm SO}({\rm level})$] on 
doubly-magic spherical nuclei.}    
\begin{tabular}{lrrrrrrrr}
\hline\hline
El.&$A$&$B\quad$&$B^{\rm exp}$&$r_{\rm c}$& $r^{\rm exp}_{\rm c}$ & 
$\Delta E_{\rm SO}$ & $\Delta E_{\rm SO}^{\rm exp}$ & (level)\\
             &     &[MeV]     & [MeV]  & [fm] & [fm] & [MeV]  & [MeV] &  \\
\hline
O  & 16& 130.48& 127.62&2.75&2.70&5.26&6.18& ($1p$)\\
Ca & 40& 347.08& 342.05&3.47&3.48&5.78&6.75& ($1d$)\\    
   & 48& 415.61& 415.99&3.51&3.47&4.75&5.30& ($1d$)\\
Ni & 56& 469.73& 483.99&3.80&--  &4.69&6.82& ($1f$)\\
Ni & 68& 593.19& 590.41&3.92&--  &--  &--  &       \\
Zr & 90& 781.26& 783.89&4.27&4.27&6.45&5.56& ($1g$)\\
Sn &100& 811.66& 824.79&4.50&--  &4.60&7.00& ($1g$)\\
   &132&1103.09&1102.85&4.73&--  &5.13&6.51& ($1h$)\\
Pb &208&1636.61&1636.43&5.50&5.50&1.88&2.02& ($2f$)\\
\hline\hline
\end{tabular}
\label{resfit}
\end{center}
\end{table}

We display in Table \ref{resfit} the SAMi results for binding energies, charge radii and proton spin-orbit splittings of all measured doubly-magic spherical nuclei. The descriptions of $B$ and $r_c$ are accurate within a 1\% and 0.6\%, respectively, and the proton spin-orbit splittings of the different single particle states with high angular momenta are accurate within 15\%. Such an accuracy will be decisive for an accurate characterization of the GTR energies \cite{bend02} and clearly improves the results obtained with SGII \cite{giai81}.   

In Fig.~\ref{gmr-gdr}, we test the performance of the SAMi interaction for the description of the strength distribution (calculated within RPA \cite{CPC}) in the cases of the GMR and GDR in ${}^{208}$Pb. The results are compared with experimental data and with the predictions of SLy5 \cite{chab98}. The operators used in the GMR and GDR cases are, respectively, $\sum_{i=1}^A r_i^2$ and $Z/A\sum_{n=1}^N r_n - N/A\sum_{p=1}^Z r_p$. The experimental centroid energy of the GMR has allowed to constrain the nuclear matter incompressibility at the value $K_\infty=240\pm20$ MeV, by means of an analysis of a large set of Skyrme interactions \cite{colo04}. Within the same spirit, the experimental data on the GDR has allowed to determine the nuclear symmetry energy at a sub-saturation density $S(\rho=0.1$ fm${}^{-3})=24.1\pm 0.8$ MeV \cite{trip08}. The SAMi interaction predicts compatible values, namely $K_\infty=245$ MeV and $S(\rho=0.1$ fm${}^{-3})= 22$ MeV. Consistently, the giant resonance centroid energy predicted by SAMi agrees well with the experimental findings: $E_{\rm c}^{\rm SAMi}({\rm GMR})= 14.48$ MeV should be compared with  $E_{\rm c}^{\rm exp.}({\rm GMR})= 14.24\pm0.11$ MeV \cite{youn99} [exhausting both almost 100\% of the Energy Weighted Sum Rule (EWSR) between $E_x=8-22$ MeV], and $E_{\rm c}^{\rm SAMi}({\rm GDR})= 13.95$ MeV should be compared with $E_{\rm c}^{\rm exp.}({\rm GDR})= 13.25\pm0.10$ MeV \cite{ryez02} (exhausting both around 95\% of the EWSR between $E_x=9-20$ MeV). 

\begin{figure}[t!]
\includegraphics[width=0.9\linewidth,clip=true]{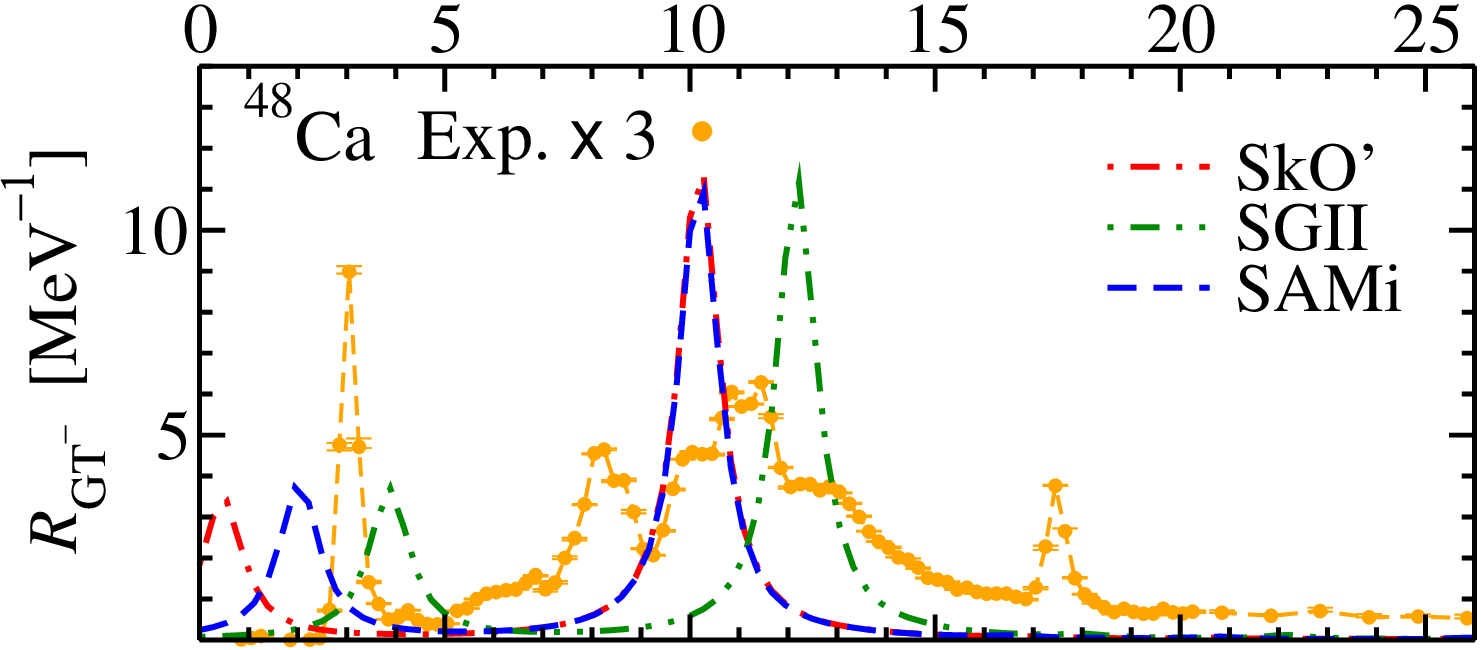}\\ 
\includegraphics[width=0.9\linewidth,clip=true]{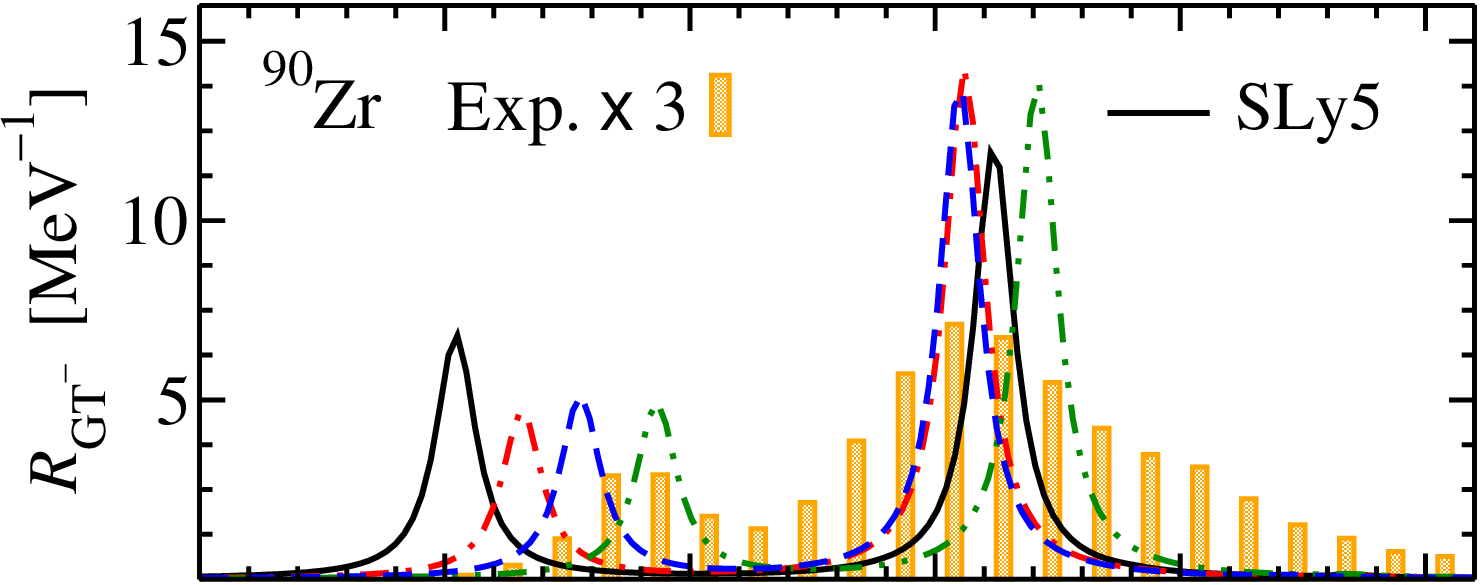}\\ 
\includegraphics[width=0.9\linewidth,clip=true]{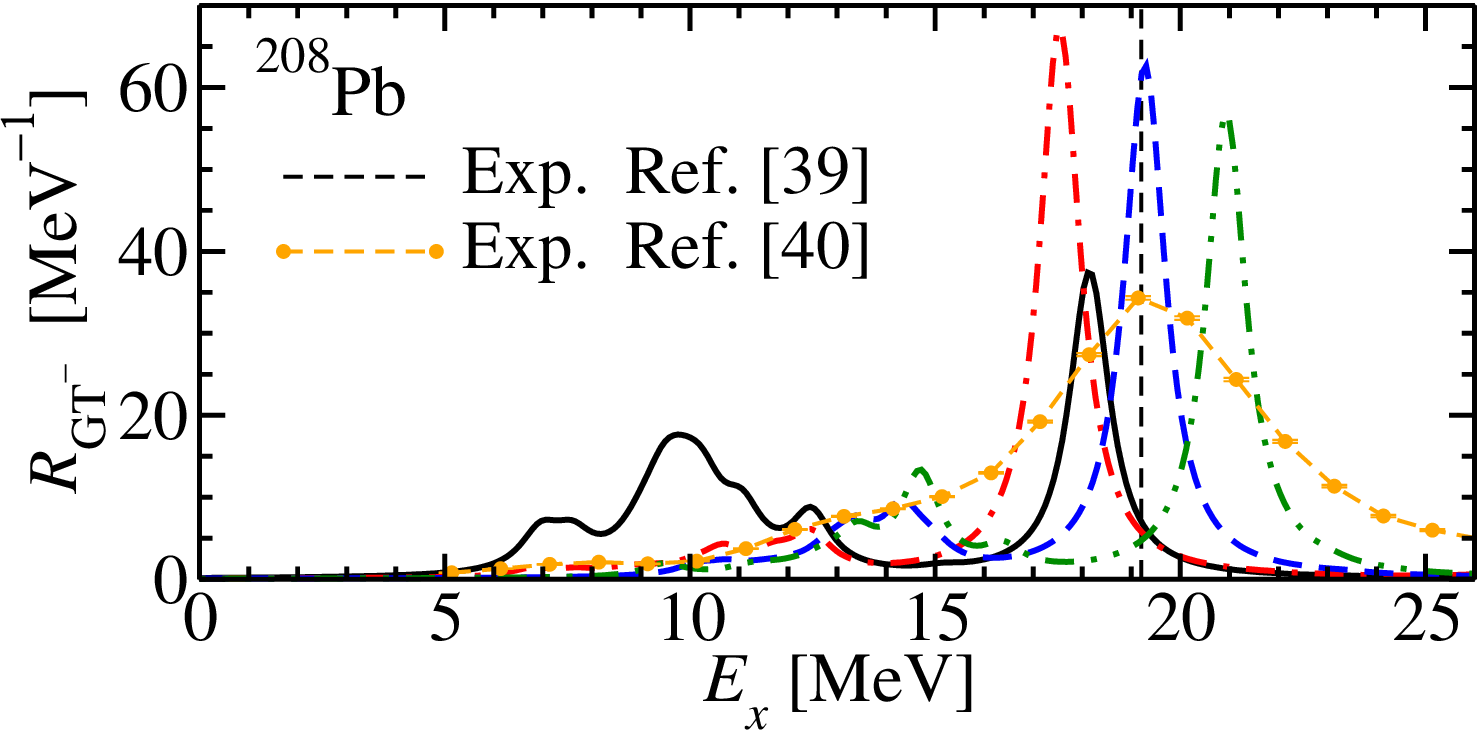}
\caption{GT strength distributions 
in ${}^{48}$Ca (upper panel), ${}^{90}$Zr (middle panel) and ${}^{208}$Pb (lower panel) as measured in the experiment \cite{yako09,waka97,kras01,akim95,waka12} and predicted by SLy5, SkO', SGII and SAMi forces.}
\label{gt}
\vspace{-0.5cm}
\end{figure}

The strength distributions of the GTR are displayed in Fig.~\ref{gt}. HF+RPA results obtained with the forces SAMi, SGII \cite{giai81}, SLy5 \cite{chab98} and SkO' \cite{rein99} are compared with experiment. In the upper panel of Fig.~\ref{gt}, we show the experimental data of Ref.~\cite{yako09} as well as the prediction of the SAMi, SGII and SkO' functionals for ${}^{48}$Ca. In this case the SLy5 result is not shown because RPA produce instabilities. The nice agreement of the SAMi prediction in the excitation energy, $E_x^{\rm exp}=10.5$ MeV and $E_x^{\rm SAMi}=10.2$ MeV for the high-energy peak and $E_x^{\rm exp}=3.0$ MeV and $E_x^{\rm SAMi}=2.0$ MeV for the low-energy peak, and the \% of the ISR exhausted by the main peak between 5 and 17 MeV, around 46\% in the experiment and 71\% in the calculation, is noticeable (in keeping with the fact that RPA does not include 2p-2h couplings). The prediction of the SAMi interaction in the case of ${}^{90}$Zr (middle panel of Fig.~\ref{gt}) is even better than in the case of $^{48}$Ca. Despite of the accuracy of SGII, SLy5 and SkO' in describing other properties of nuclei, they do not perform as well as our new proposed functional. The excitation energy and \% of the ISR exhausted by the high- and low-energy peaks in the experimental data \cite{waka97,kras01} (in the calculation done with the SAMi functional) are, respectively, $E_x^{\rm exp}=15.8\pm0.5$ MeV and 57\% ($E_x^{\rm SAMi}=15.5$ MeV and 70\%) between 12 and 30 MeV and $E_x^{\rm exp}=9.0\pm0.5$ MeV and 12\% ($E_x^{\rm SAMi}=7.8$ MeV and 27\%)  between 3 and 12 MeV. In the lower panel and with unprecedented accuracy in HF+RPA calculations, the SAMi functional perfectly reproduces the excitation energy of the experimental GTR in ${}^{208}$Pb \cite{akim95}: $E_x^{\rm exp}=19.2\pm0.2$ MeV and $E_x^{\rm SAMi}=19.3$ MeV. We also compare our results with the predictions of SGII, SLy5 and SkO' that fail in the description of the GTR in ${}^{208}$Pb. It is important to notice that, opposite to SLy5, the spin-orbit parameters ($W_0$ and $W_0^\prime$) are not fixed to be equal in the SAMi and SkO' interactions. Note also that, $G_0$ and $G_0'$ were fixed to be 0.011 and 0.503 in SGII interaction \cite{giai81} together with $K_{\infty}$=215 MeV and $J$=26.8 MeV which give reasonable descriptions for other resonances but predict the GT excitation energies in ${}^{208}$Pb at slightly higher values (21.2 MeV) -- partly due to a larger $G_0'$.

In order to assess the robustness of the SAMi functional in the description of other charge-exchange reactions, we have analyzed the Isobaric Analog Resonance (IAR) in ${}^{90}$Zr and ${}^{208}$Pb. In this nuclear excitation, modelized by the operator $\sum_{i=1}^A \tau_\pm(i)$, the nucleon transitions change the isospin of the parent quantum state. We have found that the experimental value for the IAR in both, ${}^{90}$Zr and ${}^{208}$Pb is very well reproduced by the SAMi functional -- within a 1.5\% discrepancy with respect to the experiment \cite{kras01} -- and the sum rule, $\int [R_{{\rm IA}^-}(E) - R_{{\rm IA}^+}(E)]{\rm d}E=(N-Z)$, is perfectly exhausted in our calculations.

\begin{figure}[t]
\includegraphics[width=0.9\linewidth,clip=true]{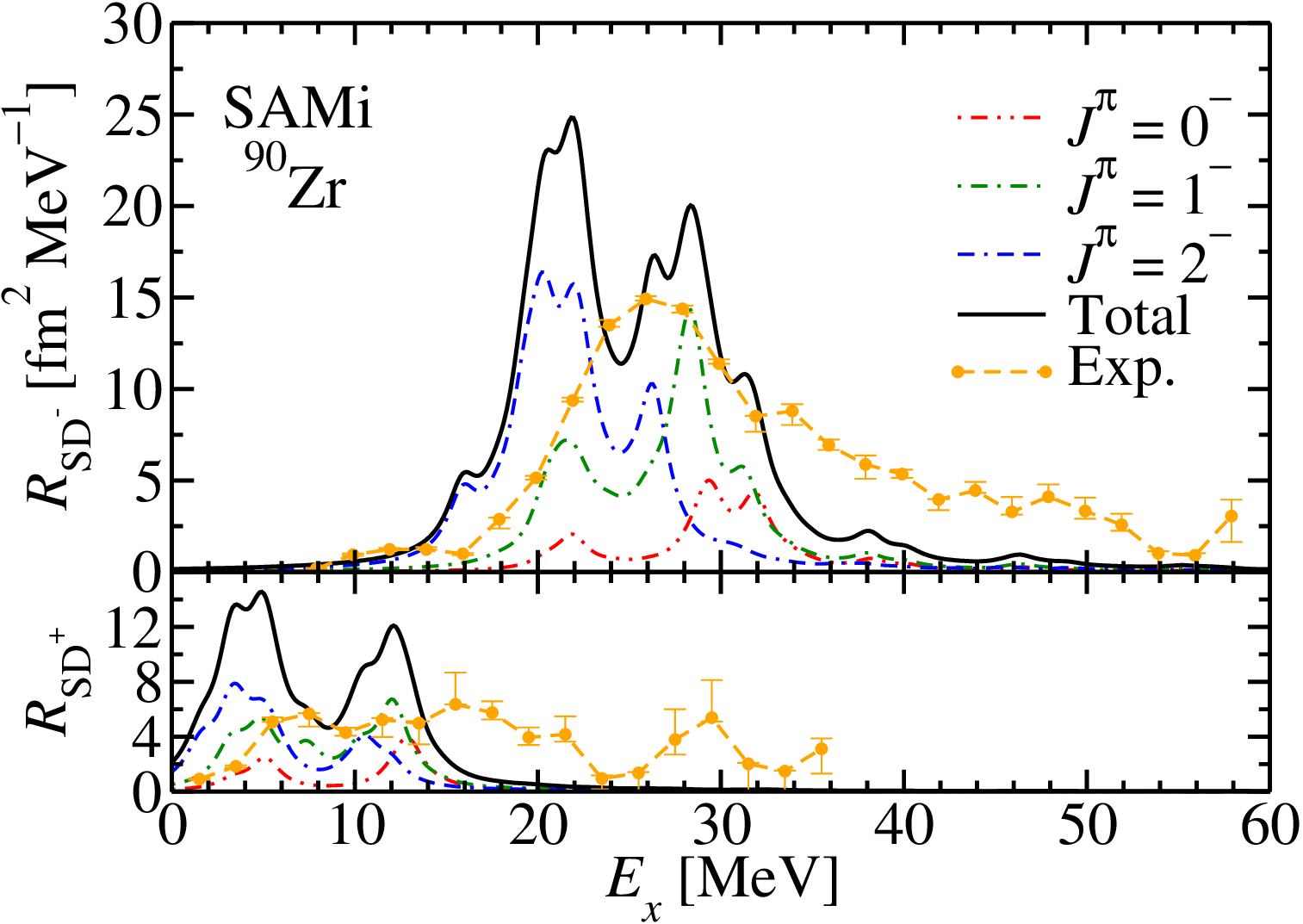}\\ 
\caption{Spin Dipole strength distributions in ${}^{90}$Zr as a function of the excitation energy $E_x$ in the $\tau_{-}$ channel (upper panel) and $\tau_{+}$ channel (lower panel) measured in the experiment \cite{yako06} and predicted by SAMi. Multipole decomposition is also shown. A Lorentzian smearing parameter equal to 2 MeV is used.}
\label{sdr-zr}
\vspace{-0.5cm}
\end{figure}

\begin{figure}[t]
\includegraphics[width=0.9\linewidth,clip=true]{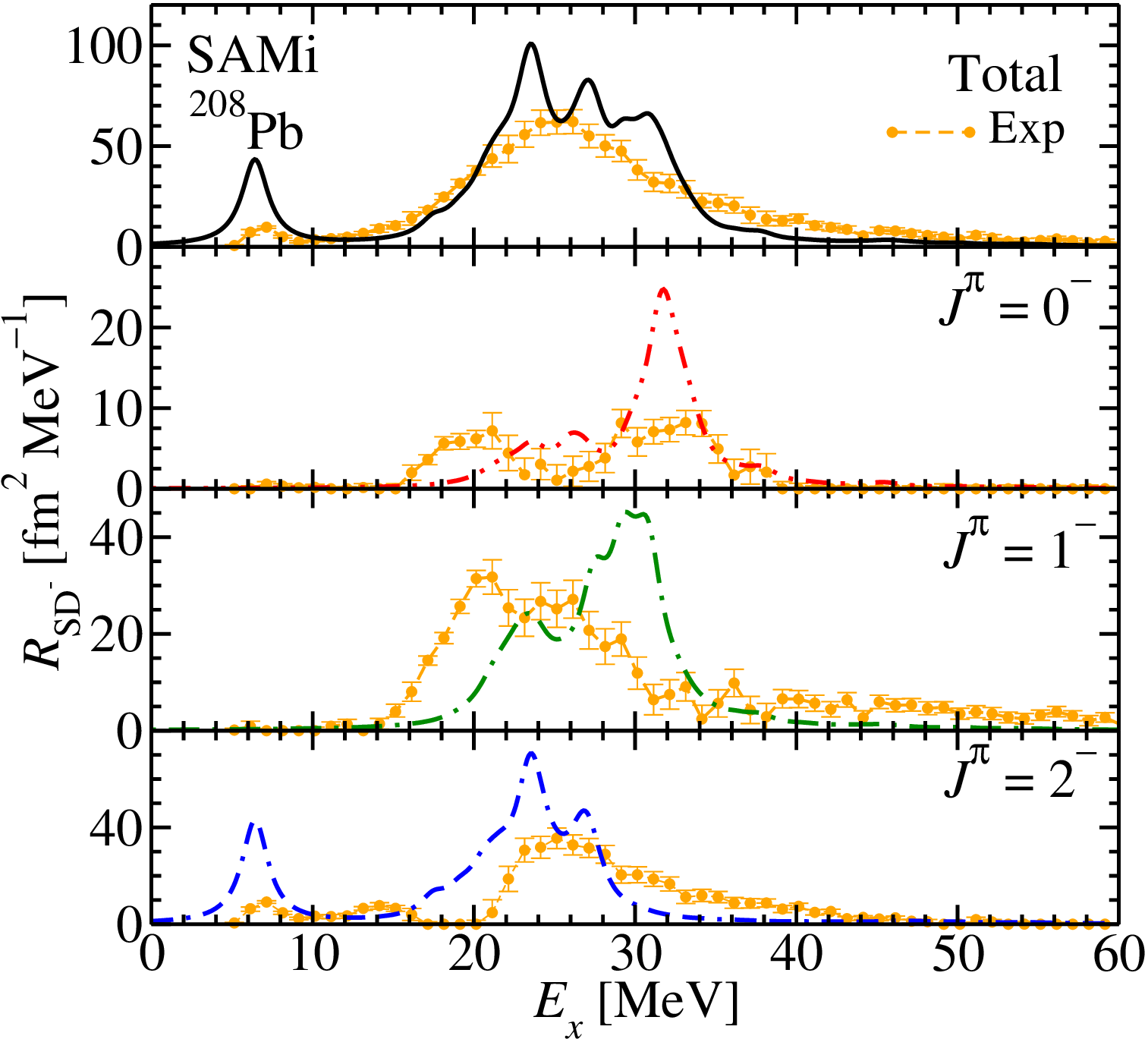}\\ 
\caption{SDR strength distributions for ${}^{208}$Pb in the $\tau_{-}$ channel from experiment \cite{waka12} and SAMi calculations. Total and multipole decomposition of the SDR strength are shown: total (upper panel), $ J^\pi = 0^-$ (middle-upper panel), $ J^\pi = 1^-$ (middle-lower panel) and $ J^\pi = 2^-$ (lower panel). A Lorentzian smearing parameter equal to 2 MeV is used.}
\label{sdr-pb}
\vspace{-0.5cm}
\end{figure}

Furthermore, experimental and theoretical studies on the Spin Dipole Resonance (SDR) have been recently revitalized due to its connection, via a sum rule, to the neutron skin thickness of nuclei ($\Delta r_{np}$) \cite{yako06} and, therefore, to the density dependence of the nuclear symmetry energy \cite{brow00,cent09}. For these reasons, we also present the SAMi predictions for this important charge-exchange excitation in ${}^{90}$Zr (Fig.~\ref{sdr-zr}) and ${}^{208}$Pb (Fig.~\ref{sdr-pb}) as compared with the experiment \cite{yako06,waka12}. The operator used for the RPA calculations is $\sum_{i=1}^A \sum_{M}\tau_\pm(i)r_i^{L}\left[Y_{L}(\hat{r}_i)\otimes\bsigma(i)\right]_{JM}$ and, as it is shown in both figures, it connects single particle states differing by a total angular momentum: $J^{\pi} = 0^-, 1^-$ and $2^-$. The sum rule, $\int [R_{{\rm SD}^-}(E) - R_{{\rm SD}^+}(E)]{\rm d}E=\frac{9}{4\pi}(N\langle r_n^2\rangle-Z\langle r_p^2\rangle)$, is completely exhausted in our calculations, 99.99\% in the case of ${}^{90}$Zr and 100\% in the case of ${}^{208}$Pb. The experimental (calculated) value for the sum rule is $148\pm12$ fm${}^{2}$ \cite{yako06} ($150$ fm${}^{2}$) for the case of ${}^{90}$Zr. The total and multipole decomposition of the experimental \cite{waka12} (calculated) value for the integral of $R_{{\rm SD}^-}$ in the case of ${}^{208}$Pb are 1004$^{+24}_{-23}$ fm${}^{2}$ (1224 fm${}^{2}$), 107$^{+8}_{-7}$ fm${}^{2}$ (158 fm${}^{2}$) for $J^{\pi} = 0^-$, 450$^{+16}_{-15}$ fm${}^{2}$ (423 fm${}^{2}$) for $J^{\pi} = 1^-$ and 447$^{+16}_{-15}$ fm${}^{2}$ (643 fm${}^{2}$) for $J^{\pi} = 2^-$. Finally, it is noticeable the overall agreement between experiment and SAMi predictions when the strength distributions as a function of the excitation energies shown in Figs.~\ref{sdr-zr} and \ref{sdr-pb} are compared.    

The neutron skin thickness of medium and heavy nuclei is known to be strongly correlated with the isospin properties of the nuclear effective interaction \cite{brow00,cent09}. A recent study \cite{tsan12} shows that the $\Delta r_{np}$ in ${}^{208}$Pb derived from different hadronic probes agrees in a value of 0.18$\pm$0.03 fm. The SAMi interaction predicts 0.15 fm, compatible within the estimated error bars. In addition, a recent theoretical study \cite{piek12} has allowed to predict $0.17\pm0.02$ fm. Recently, the PREx collaboration has reported a value of $0.33_{-0.18}^{+0.16}$ fm for the same observable measured via parity violating elastic electron scattering \cite{prex12,roca11}. If this value is confirmed with high accuracy, a deep revision of current nuclear models will be necessary. In the case of ${}^{90}$Zr, an analysis of the charge exchange Spin-Dipole resonance has allowed to extract $\Delta r_{np}({}^{90}{\rm Zr})=0.07\pm0.04$ fm \cite{yako06}, in perfect accordance with our predicted value of 0.07 fm: an additional proof of the improvement in the description of the spin and isospin channels of the nuclear effective interaction provided by SAMi. Finally, the neutron skin in ${}^{48}$Ca is predicted by our model to be 0.17 fm very close to the theoretical value $0.176\pm0.018$ fm reported in \cite{piek12}. 

In summary, we have successfully determined a new Skyrme energy density functional which accounts for the most relevant quantities in order to improve the description of charge-exchange nuclear resonances, i.e., the hierarchy and positive values of the spin and spin-isospin Landau-Migdal parameters and the proton spin-orbit splittings of different high angular momenta single-particle levels. As a proof, the GTR in $^{48}$Ca and the GTR, IAR and SDR in $^{90}$Zr and $^{208}$Pb are predicted with high accuracy by SAMi without deteriorating the description of other nuclear observables and, therefore, promising its wide applicability in nuclear physics and astrophysics.   

We are very grateful to T. Wakasa, H. Sakai and K. Yako for providing us with the experimental data on the presented charge-exchange resonances for ${}^{48}$Ca, ${}^{90}$Zr and ${}^{208}$Pb and to M. Baldo, D. Gambacurta, and I. Vida\~na for the state-of-the-art BHF calculations. The support of the Italian Research Project ``Many-body theory of nuclear systems and implications on the physics of neutron stars'' (PRIN 2008) is acknowledged.


%
\end{document}